\documentclass[ apl,reprint,superscriptaddress]{revtex4-1}
\usepackage[english]{babel}
\usepackage{graphicx}
\usepackage{multirow}
\usepackage{color}
\usepackage{amsmath,amssymb}
\usepackage[latin1]{inputenc}

\begin{document}

\title{Microscopic mechanism of the giant magnetocaloric effect in MnCoGe alloys probed by XMCD}%

\begin{abstract}
One important aspect of the magneto-structural transition in MnCoGe and related materials is the reduction in saturation magnetization from the orthorhombic to the hexagonal phase. Here, by combining an element specific magnetization probe such as x-ray magnetic circular dichroism and band structure calculations, we show that the magnetic moment instability between orthorhombic and hexagonal structures originates from a reduction in the Mn sub-lattice magnetization. The consequences of the moment instability for the magnetocaloric effect are discussed.

\end{abstract}
\author{F. Guillou} 
\email{francois.guillou@esrf.fr}
\affiliation{ESRF, 71 Av. des Martyrs, F-38043 Grenoble Cedex 09, France}
\author{F. Wilhelm}
\affiliation{ESRF, 71 Av. des Martyrs, F-38043 Grenoble Cedex 09, France}
\author{O. Tegus}
\affiliation{Inner Mongolia Key Laboratory for Physics and Chemistry of Functional Materials, Hohhot 010022, China}
\author{A. Rogalev}
\affiliation{ESRF, 71 Av. des Martyrs, F-38043 Grenoble Cedex 09, France}
\date{\today}%
\maketitle

Among the emerging applications of functional magnetic materials, magnetic refrigeration is particularly appealing due to its energetic and environmental advantages compared to conventional cooling methods. It is based on the magnetocaloric effect (MCE), the temperature change of a material induced by a magnetic field. A giant-MCE is observed in materials exhibiting first-order magnetic transition (FOMT), like when the magnetic field triggers a change in crystal structure. Understanding the relation between structure and magnetic properties is thus of high applicative and fundamental importance as it paves the way for further improvement of the MCE. MnCoGe-based compounds have recently attracted great interest. MnCoGe shows a second-order type transition to a ferromagnetic state with a Curie temperature ($T_{C}$) close to 290 K. It also displays a structural transition from orthorhombic (ortho.) TiNiSi-type to hexagonal (hexa.) Ni$_{2}$In-type at $T$= 640 K. A modest chemical pressure or application of an external pressure allow the structural transition to be tuned over a large temperature range. This offers a possibility to make structural and magnetic transitions to coincide at a single ferromagnetic (FM) to paramagnetic (PM) magneto-structural transition.\cite{JOH75, LIN06, TRUN10}  This transition displays an intense latent heat important for magnetic refrigeration,\cite{TRUN10,LIN06,Liu2012642,SAM12,CHOU14} and a giant volume change which can be used for negative thermal expansion or magnetostrictive applications.\cite{ZHAO15} Whereas the latent heat is particularly large in MnCoGe-compounds, turning it into high temperature changes is however requiring an intense magnetic field. Making the interplay between the different degrees of freedom at FOMT more efficient is actually a key issue in all giant-MCE materials. At magneto-structural transitions, it primarily concerns the respective roles of magnetic and structural changes and calls for a detailed understanding of the microscopic mechanisms involved.

Most of the studies on the MnCoGe-based materials are focused on the search for chemical compositions with large MCE. Using this trial and error approach, more than thirty material systems have been reported so far. It remains however a heuristic method. To optimize the quest for more efficient materials, it is highly desirable to understand the transition, the way it develops and the microscopic mechanisms at play. Only few studies have addressed the coupling between structural and magnetic transitions in MnCoGe.\cite{ANZ78,Caron11,GschneidnerJr2012572} Depending on the thermal treatment, MnCoGe can be obtained either in the orthorhombic or in the hexagonal structure with saturation magnetization of 3.73 $\mu_{B}$/f.u. and 2.80 $\mu_{B}$/f.u., respectively.\cite{Kap90} This reduction of magnetization in MnCoGe-based compounds is often disregarded, while it presumably plays an important role. For giant-MCE materials with FM ortho. to PM hexa. transition, it means that the MCE not only originates from a change of crystal structure, but is also due to a loss of local magnetic moments. Unfortunately, the microscopic mechanism remains unclear, in particular in respect to which element(s) is (are) the most affected. One could expect that in such metallic alloys both transition-metals exhibit a similar evolution. A neutron diffraction study indicates that the magnetization reduction is associated with changes in local magnetic moments of both Mn and Co atoms, from 3.16 $\mu_{B}$ to 2.3 $\mu_{B}$ and 0.7 $\mu_{B}$ to 0.1 $\mu_{B}$, respectively. On the other hand, {\it ab-initio} calculations suggested that either only the Co magnetic moments or only the Mn ones evolve at the transition.\cite{Kap90,RUS08}

The aim of the present work is to elucidate the role of each element in the FOMT and giant-MCE by using an element specific magnetic probe. MnCoGe with B interstitial addition is one archetypical example of giant-MCE in this system and offers the possibility to study ortho. and hexa. structures.\cite{TRUN10} XMCD spectra at the $K$-edges of Mn, Co and Ge have been recorded to get element specific magnetic properties. This choice is motivated by the pulverization of bulk samples across the transition which makes the preparation of clean surfaces impossible. The large penetration depth of hard x-rays allows truly bulk properties to be measured. The experimental results were compared with theoretical simulations to relate the XMCD amplitude to the predicted changes in magnetic moments.

Polycrystalline MnCoGeB$_{0.04}$ samples were synthesized by arc-melting of high-purity Mn, Co, Ge and B elemental precursors. The ingot was then split and annealed in evacuated quartz tube (1000$^{o}$C). Two different samples were obtained by varying the annealing conditions, slow cooling or quenching, for samples referred to as A and B, respectively. The magnetization measurements were carried out using MPMS SQUID magnetometer. The XMCD experiments were performed at the ESRF beamline ID12 on the experimental station equipped with a superconducting magnet allowing magnetic field up to 17 T.\cite{ID12beamline} For these experiments,  MnCoGeB$_{0.04}$ powdered samples were mixed with carbon and pressed into pellets. The x-ray absorption spectra were recorded in total fluorescence yield detection mode with Si photo-diode in backscattering geometry. The x-ray absorption and consequently XMCD spectra were corrected from self-absorption effects by assuming  semi-infinite  samples, taking  into account  the  various  contributions  (fluorescence  of  subshells and matrix as well as coherent and incoherent scattering), the angle of incidence of the x-ray beam, and the solid angle of the detector.

\begin{figure}
     \centering
     \includegraphics[trim = 7mm 8.5mm 2mm 17.5mm, clip,width=0.5\textwidth]{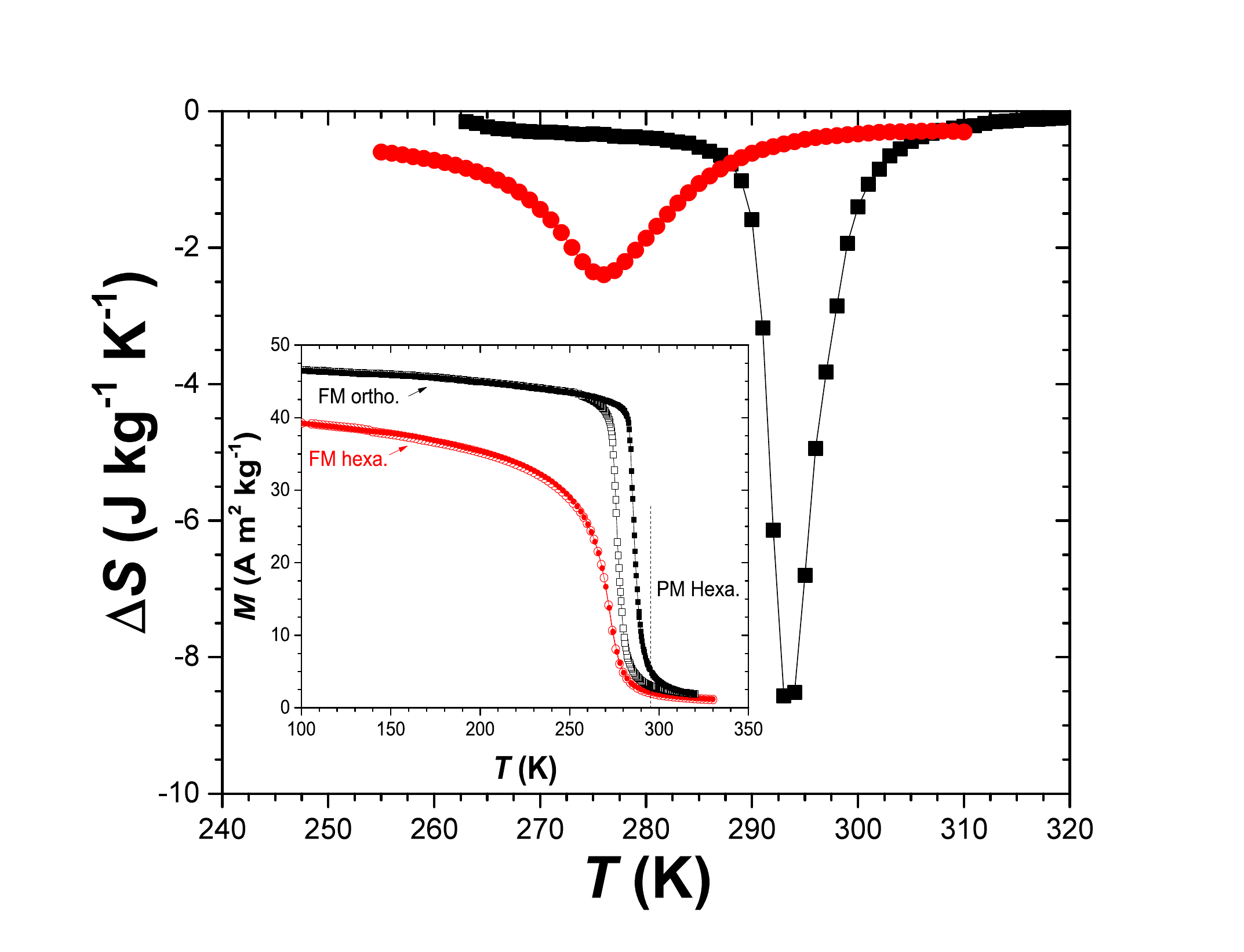}
     \caption{ Magnetocaloric entropy change curves for 1 T calculated from $M(T)$ measurements upon warming. In inset, magnetization as a function of the temperature at 0.1 T for MnCoGeB$_{0.04}$ samples: A (squares) and B (circles) upon warming (filled symbols) and cooling (open symbols).}
     \label{fig1}
\end{figure}

The magnetocaloric entropy change and macroscopic magnetization curves as a function of the temperature are shown in Fig.\ref{fig1} for the two prototypical samples. Sample A clearly shows the magneto-structural transition from FM ortho. to PM hexa. at 290 K. On the other hand, the sample B, which remains hexagonal in the temperature range from 100 K to 325 K, does not show any thermal hysteresis at $T_{C}$ = 275~K and the broad magnetization change is typical of a continuous transition. The $\Delta S$ maximum is thrice larger in A than in B, in line with previous reports in MnCoGe-based materials.\cite{TRUN10,LIN06} This highlights how crucial the FOMT is for the MCE performances. From magnetization measurements up to 5 T at 150 K, saturation magnetization of about 3.74~$\mu_{B}$/f.u. and 2.98~$\mu_{B}$/f.u. were obtained for samples A and B, respectively. These values are typical for MnCoGe compounds and are similar to that of the parent composition in ortho. and hexa. forms.\cite{Kap90}

\begin{table}
\caption{\label{table1} Magnetic moments derived from band structures calculations for MnCoGe.}
\begin{ruledtabular}
\begin{tabular}{ccc}
\multirow{2} {*} {Atoms}         &  TiNiSi (FM ortho.)       & Ni$_{2}$In (FM hexa.)          \\ 
							& ($\mu_{B}$)	& ($\mu_{B}$)	 \\ 
\hline
Mn        		&  3.335 		& 2.800    		 \\ 
Co          &0.449		     & 0.442		   \\ 
Ge      &-0.095			&-0.090		      \\ 
 \hline
Total  (calc.)      & 3.689			& 3.152        \\ 
 \hline
Total  (exp.)      & 3.74			& 2.98        \\ 
\end{tabular}
\end{ruledtabular}
\end{table}

The isotropic XAS spectra were recorded for both samples above (300 K) and below the Curie temperature (150 K). X-ray absorption spectra are governed by the selection rules, and, at the $K$-edge of $3d$ transition-metals or metalloids they are reflecting mainly the $4p$ unoccupied density of states. The changes in the densities of states between the ortho. and hexa. structures are echoed in the XAS spectra at the $K$- edges. The XMCD spectra were measured at 150 K and 5 T, i.e. at magnetic saturation for the FM ortho. (A) and FM hexa. phases (B), respectively. XMCD spectra at the $K$-edge reflect the orbital polarization density of the conduction $p$ states, connected to the spin polarization though the spin-orbit interaction.\cite{xmcd, xmcd2} For the transition metals, the delocalized $4p$ states are coupled with the spin-polarized $3d$ states by exchange interaction, so that the $K$-edge XMCD of transition metals is a fingerprint of the local moment carried by the exited atom and, for a given material, provides a mean for measuring the evolution of the magnetization in an element selective manner.

\begin{figure}
     \centering
     \includegraphics[trim = 3mm 10.7mm 0mm 13.5mm, clip,width=0.5\textwidth]{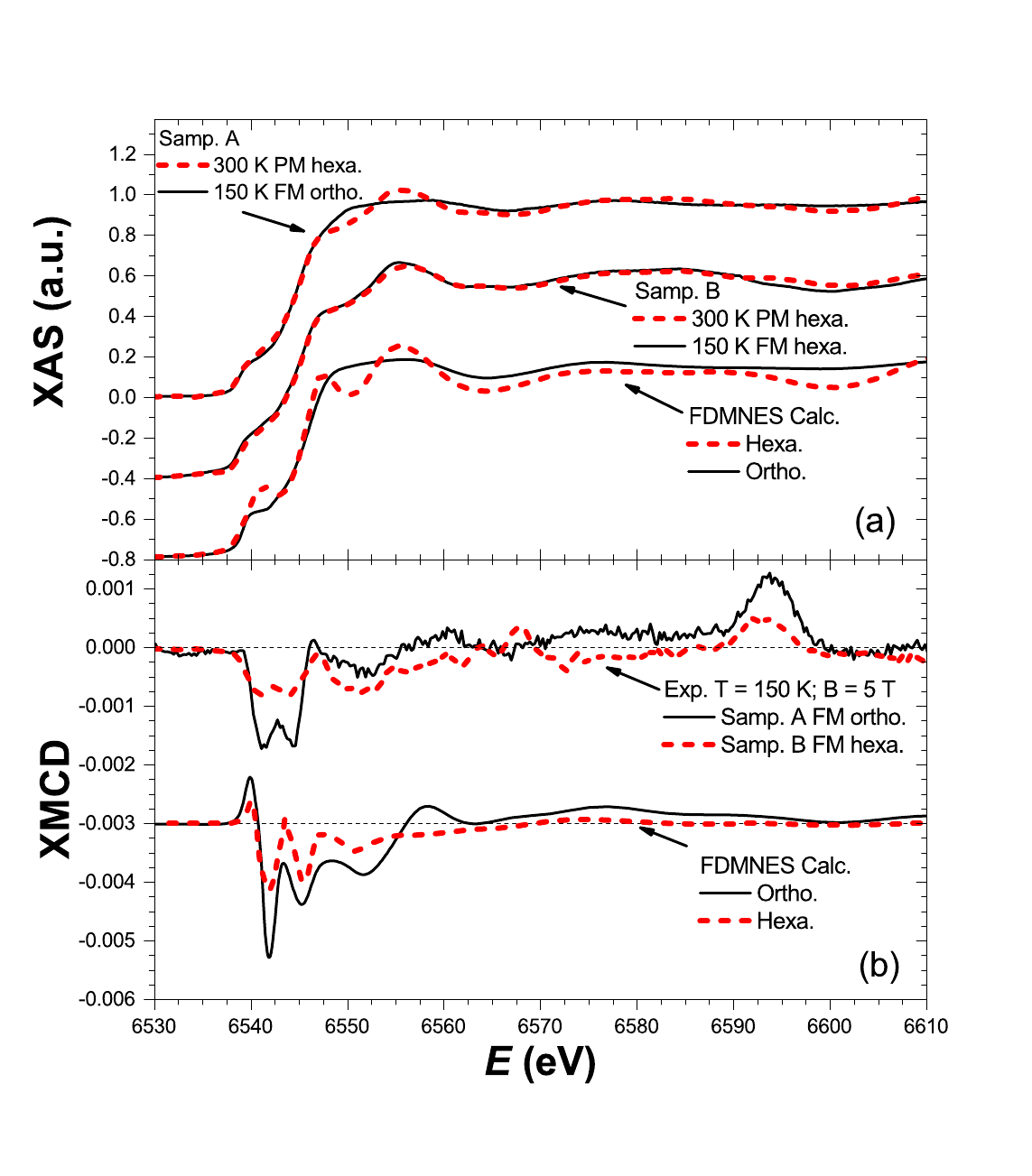}
     \caption{Mn $K$-edge XAS (a) and normalized XMCD (b) spectra: comparison between experimental spectra (top curves) and FDMNES calculations (bottom curves)}
     \label{fig2}
\end{figure}

\begin{figure}
     \centering
     \includegraphics[trim = 3mm 11.7mm 0mm 13.5mm, clip,width=0.5\textwidth]{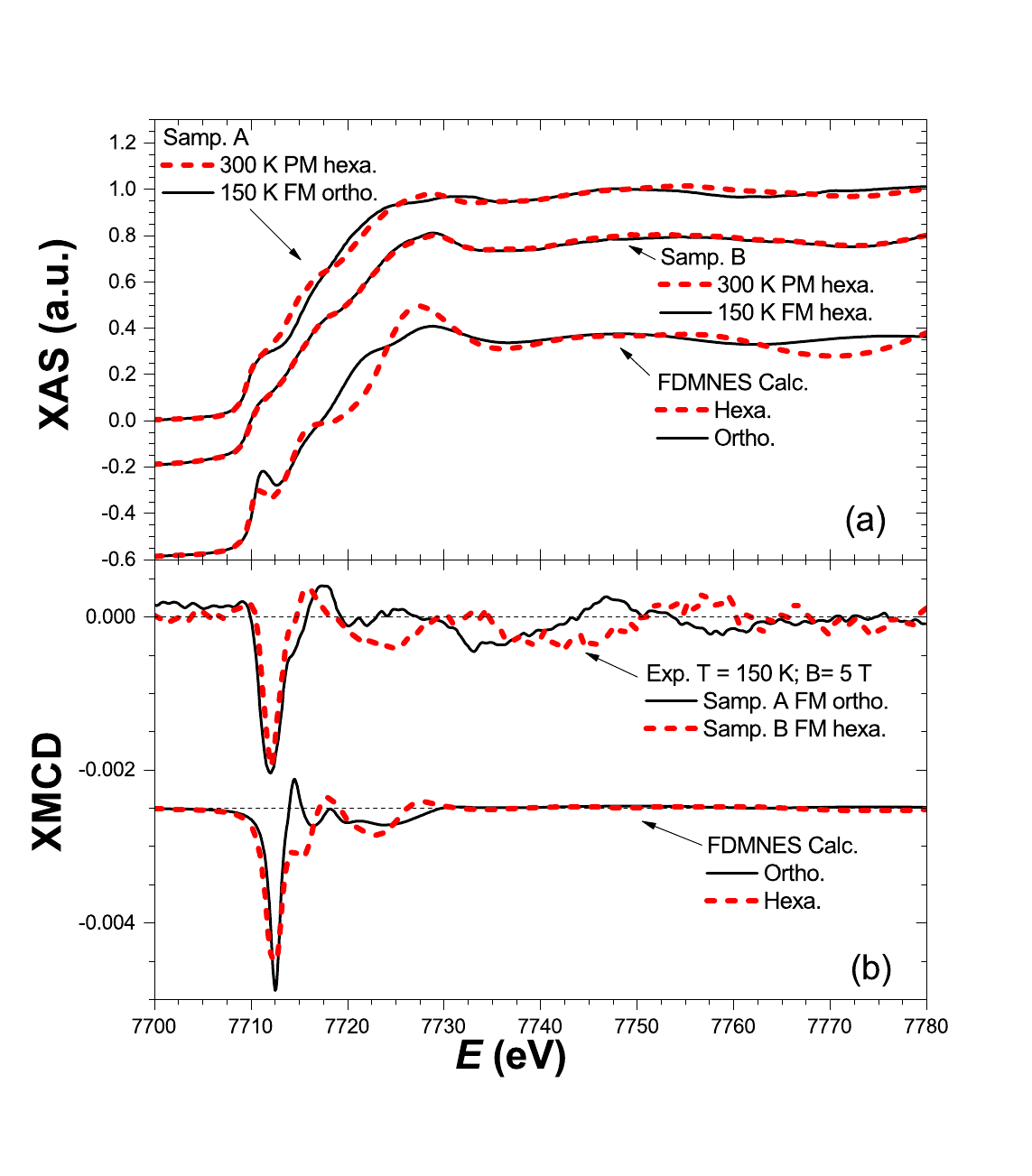}
     \caption{Co $K$-edge XAS (a) and normalized XMCD (b) spectra: comparison between experimental spectra (top curves) and FDMNES calculations (bottom curves)}
     \label{fig3}
\end{figure}

To guide the analysis of the x-ray spectroscopy experiments, in particular to get an estimate of the local magnetic moments and then to simulate the corresponding XAS and XMCD spectra, calculations were performed using two different codes. The magnetic moments determined by self-consistent relativistic spin-polarized band structure calculations using the linear muffin-tin orbitals method as implemented in the PY-LMTO code are given in Table \ref{table1}.\cite{YAR} We used the experimental lattice parameters and atomic positions of MnCoGe in the ortho. FM and hexa. FM phases.\cite{note3} The calculated total magnetization for both FM ortho. and FM hexa. phases are in satisfactory agreement with the experimental values. Comparing the individual contributions, it appears that the total magnetization difference between the two phases almost entirely originates from the Mn $3d$ magnetic moments. It is worth to underline that not only $3d$ magnetic moments of Mn but also the $4p$ orbital moments were found to show a pronounced evolution from the ortho. to the hexa. structure. The magnetic moments are in fact reflecting significant changes observed in the partial densities of states of the three elements. However, the most significant modification occurs on the $3d$ states of Mn with a band broadening from ortho. to hexa. related to the large changes in Mn-Mn interatomic distances, which may explain the larger Mn moments in the ortho. phase.\cite{Kap90}  It is also interesting to note that Ge atoms, usually considered to be non-magnetic, are predicted to acquire, via hybridization with $3d$ states of transition-metals, a finite magnetization of similar amplitude in both ortho. and hexa. structures. Even though the total magnetization of Ge atoms is opposite to that of transition-metals, the calculated $4p$ orbital moments of Ge were found to be positive and parallel to that of Mn and Co. The results of these calculations lead to a picture opposite to that drawn by KKR-LSD calculations and neutron diffraction experiments.\cite{Kap90}

In order to simulate the XAS and XMCD spectra and their evolutions across the structural transition, the FDMNES code has been used.\cite{joly,note3} It has to be stressed that these calculations were performed with the only aim to understand qualitatively the spectral changes and expected XMCD amplitudes for the two phases. As a consequence, the calculated spectra shown hereafter correspond to raw outputs, which were on purpose not refined to match the experimental data. It appears that the results of the FDMNES code is highly sensitive to the local electron and spin densities of the absorbing atoms given in input. These latter values were thus taken from the results of the relativistic band structure calculations described above. In particular, the initial spin-up and spin-down populations of transition-metals were chosen to reproduce the magnetic moments given in Table \ref{table1}.

The Mn $K$-edge spectra for samples A and B are presented in Fig.\ref{fig2}. For sample A, crossing the magneto-structural transition leads to significant changes in the x-ray absorption spectra. The overall spectral evolution is qualitatively reproduced by the FDMNES simulations. On the other hand, the XAS spectra for sample B do not show changes across the paramagnetic to ferromagnetic transition and are almost identical to that of sample A in the PM hexa. state. Below the Curie temperature, a significant difference in the amplitude of the XMCD for samples A and B is observed at the $K$-edge of Mn. It is about twice lower in FM hexa. than in FM ortho. state. This reduction of the $4p$ polarization strongly suggests a reduction of the overall magnetic moments carried by Mn atoms. For the FM ortho. state, a fair agreement between the simulation and experimental XMCD spectra is observed. The three main negative features are reproduced. For the FM hexa. structure, a reduction in the difference between $3d$ spin up-down populations of Mn is required to reproduce the reduction in XMCD signal in the simulations. Note that a relatively large positive peak is observed on the experimental XMCD spectra at photon energies about 55 eV above the edge. This feature originates from multi-electronic excitations,\cite{MME} and it also shows a strong reduction from the ortho. to hexa. structure. The reduction in XMCD intensity between sample A and B at the Mn $K$-edge demonstrates a reduction of the local magnetic moments carried by Mn atoms as predicted by the band-structure calculations.

XAS and XMCD spectra at the Co $K$-edge of samples A and B are shown in Fig.\ref{fig3}. For the sample A, the changes in XAS spectra across the FOMT are well noticeable: only one hump in the edge for FM ortho., while a second feature appears in the PM hexa. state. It turns out that the unrefined simulations already lead to a fair qualitative agreement with the experimental XAS spectra. In contrast to the Mn $K$-edge XMCD spectra, the intensity of the main XMCD peak at the Co $K$-edge is practically constant between the ortho. and hexa. phases. This is in full agreement with the results of the band structure calculations summarized in Table \ref{table1} where no differences are predicted for the Co magnetic moments. Note, that the close similarity in XMCD intensity for the two structures is also seen on the simulated curves.

\begin{figure}
     \centering
     \includegraphics[trim = 3mm 11.4mm 0mm 12.7mm, clip, width=0.5\textwidth]{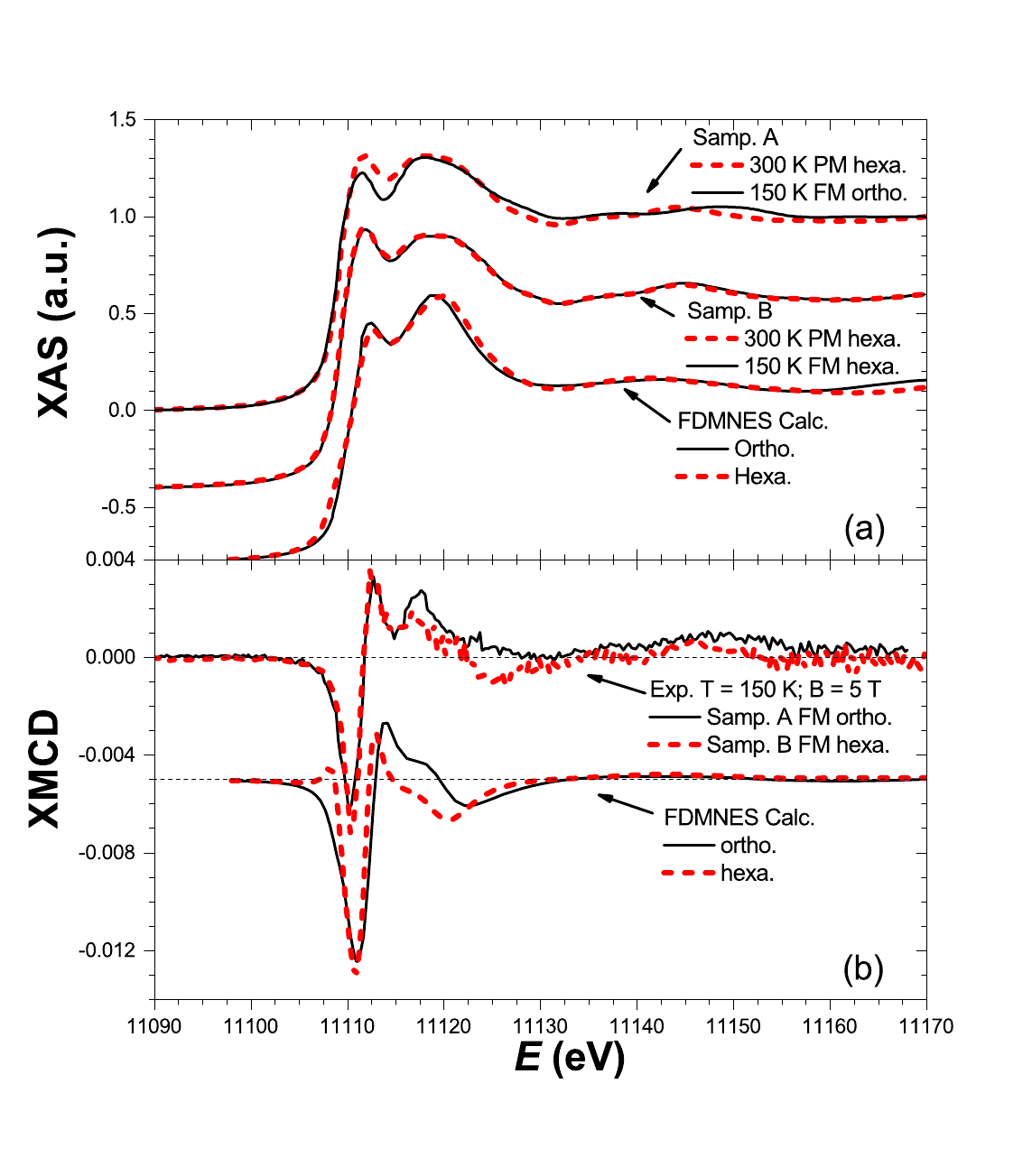}
     \caption{Ge $K$-edge XAS (a) and normalized XMCD (b) spectra: comparison between experimental spectra (top curves) and FDMNES calculations (bottom curves)}
     \label{fig4}
\end{figure}

XAS and XMCD spectra at the Ge $K$-edge are presented in Fig.\ref{fig4}. The FM to PM transition corresponds to an evolution of the XANES spectra for sample A only. XANES and XMCD spectral shapes for each structure are well consistent with those expected from the simulations. The experimental XMCD spectra indicate that the $4p$ orbital moments of Ge are parallel to the $4p$ orbital moments of the transition metals, and Ge is developing a magnetic polarization of similar amplitude in both ortho. and hexa. structures. Despite its induced character, the polarization of Ge does not seem to be affected by the reduction in magnetization of Mn, which is in line with the results shown in Table  \ref{table1}. This observation points out a stronger hybridization of Ge $4p$ states with Co than with Mn. This is no unexpected as the Ge-Co distances (2.3 \AA) are much shorter than Ge-Mn distances (2.7 \AA).\cite{Kap90}

The results of XMCD experiments and band-structure calculations demonstrate that the difference in saturation magnetization between ortho. and hexa. structures in MnCoGe compounds is solely due to the reduction in Mn local moments. For compositions with ortho. FM - hexa. PM transition, the instability of the Mn local moments is the key point to describe their giant-MCE at a microscopic level. The contribution of the Mn moment reduction to the MCE could be twofold. On one hand, the reduction in Mn moments will amplify the difference in magnetization $\Delta M$ across the transition, which is necessarily positive in terms of performance. On the other hand, in the simplified framework of the magnetic entropy in the paramagnetic state defined by the total angular momentum $J$ from the formula $S_{mag}=N_{Av}k_{B}ln(2J+1)$, the reduction in magnetic moment in the hexa. structure lowers the magnetic entropy. This would actually reduce the entropy difference between the two phases, which would in turn be negative for the MCE. One can speculate that the effect of the Mn moments reduction remains in average positive for the giant-MCE as the reduction in $\Delta M$ will act on the sensitivity of the transition to the magnetic field, which is connected to the total latent heat through the Clausius-Clapeyron formalism. The reduction in magnetic entropy most probably remains a second-order effect in such a system for which the structural contribution to the total entropy change is dominant.\cite{GschneidnerJr2012572}

To conclude, the combined experimental and theoretical analysis of XAS and XMCD spectra at the $K$-edges of all elements in MnCoGe ortho. and hexa. structures revealed a reduction in the Mn magnetic moments. At the FOMT, while all elements show an evolution of their x-ray absorption spectra and thus of their local electronic structure, a reduction in XMCD signal is observed only for Mn. This instability of the Mn local moments is an important aspect of this class of materials. It most likely is the main driving force of the FOMT and giant-magnetocaloric effect. In practical terms for the search of MCE materials, this study for instance implies that chemical substitutions should not be made at the expense of Mn. Paying more attention to maximize the Mn moment instability would improve the sensitivity of the magneto-structural transition to the magnetic field, which would in turn enhance the MCE. Shedding into light this phenomenon by both experimental and theoretical means opens new routes for designing giant-MCE materials in this system,  and it paves the way for advanced modelization of the transition.

The authors would like to thank: A. Yaresko for his help with the PY-LMTO code, V. Hardy for providing magnetometer time, and P. Voisin for the technical support. 

\bibliography{L15-10552_Guillou_revision}

\begin{thebibliography}{19}
\bibitem [{not()}]{note3}%
  \BibitemOpen
  \href@noop {} {\ }\bibinfo {note} {The LMTO or FDMNES calculations were
  performed without considering B atoms and with the following experimental
  structures, for the ortho. phase: $a$= 5.959 \AA, $b$= 3.814 \AA, $c$= 7.032
  \AA, x$_{Co}$ =0.155, z$_{Co}$ = 0.6, x$_{Mn}$= 0.02, z$_{Mn}$= 0.19,
  x$_{Ge}$= 0.78, z$_{Ge}$= 0.61 ; and for the hexa. phase: a = 4.034 \AA, c =
  5.241 \AA. For the FDMNES simulations, the following parameters were used:
  cluster of 6 \AA, muffin-tin type potential, only the dipolar transitions are
  considered, as-implemented automatic convolution, no adjustments of the
  parameters of the convolution, no additional convolution for the experimental
  energy resolution.}\BibitemShut {Stop}%
\end{thebibliography}

\end{document}